\documentclass[sigconf, nonacm]{acmart}
\usepackage{graphicx}
\usepackage{fontawesome}
\usepackage{float}
\usepackage{graphicx} 
\usepackage{subfigure}
\usepackage{algorithm}
\usepackage{algorithmic}
\usepackage{amsmath}

\AtBeginDocument{%
  \providecommand\BibTeX{{%
    \normalfont B\kern-0.5em{\scshape i\kern-0.25em b}\kern-0.8em\TeX}}}

\acmConference[ADVANCE '26]{International Workshop on ADVANCEs in ICT Infrastructures and Services}{March 25th-27th, 2026}{Florianópolis - Brazil}
\acmBooktitle{ADVANCE '26: Proceedings of the International Workshop on ADVANCEs in ICT Infrastructures and Services, March 25th-27th, 2026, Florianópolis, Brazil} 

\begin{document}





\title{Dynamic Map-based Data-Centric Approach for Tourism and Cultural Heritage Preservation Digital Twins}

\author{João Spínola Falcão}
\orcid{0000-0001-7051-7396}
\email{joaofalcao2004@gmail.com}
\affiliation{
  \institution{Universidade Salvador (UNIFACS)}
  \city{Salvador}
  \country{Brazil}
}

\author{João G. Perrone Hohlenwerger}
\email{joaoperrone1831@gmail.com}
\affiliation{
  \institution{Universidade Salvador (UNIFACS)}
  \city{Salvador}
  \country{Brazil}
}

\author{Daniel C. Santos}
\email{dann.costasantos@gmail.com}
\affiliation{
  \institution{Universidade Salvador (UNIFACS)}
  \city{Salvador}
  \country{Brazil}
}

\author{Lucas Almeida de Sousa }
\orcid{0000-0003-1310-9366}
\email{lucas.ads@fieb.org.br}
\affiliation{%
  \institution{Universidade Salvador (UNIFACS)}
  \city{Salvador}
  \country{Brazil}
}


\author{Nazim Agoulmine}
\orcid{}
\email{nagoulmine@gmail.com}
\affiliation{%
  \institution{Université Paris-Saclay - Évry}
  \city{Évry}
  \country{France}
}
\author{Joberto S. B. Martins}
\authornote{All authors contributed equally to this research.}
\orcid{0000-0003-1310-9366}
\email{joberto.martins@gmail.com}
\affiliation{
  \institution{Universidade Salvador (UNIFACS)}
  \streetaddress{Av. ACM 1133}
  \city{Salvador}
  \country{Brazil}}


\renewcommand{\shortauthors}{Spínola and Martins}

\begin{abstract}
 
Tourism is an essential and growing economic activity worldwide, bringing benefits such as job creation, revenue generation, and tax revenue, and driving economic prosperity. Tourism activity may also have negative impacts on cities, such as overtourism and pressure on housing and real estate, which are increasingly recognized as problems communities must address. Cultural heritage is an essential asset of cities and countries that must be preserved. Cultural heritage, as an asset, is commonly explored through tourism activities and may have negative impacts, including physical degradation, commodification, and loss of authenticity, among others. Tourism and cultural heritage management are common elements of smart city digital transformation strategies, in which the well-being of citizens and the maintenance of public assets are among the goals. A digital twin is a data-driven virtual representation of a physical object, system, or environment. It typically integrates real-time data and computational models to simulate, monitor, analyze, and predict the behavior and performance of the represented entity. However, although digital twin technology has been widely adopted in manufacturing, Industry 4.0, and urban planning for smart cities, there remains a gap in specialized digital twins for tourism and cultural heritage management. This paper proposes a QGIS-based, data-centric approach to digital twin frameworks that supports the management, development, and deployment of tourism and cultural heritage services and applications in smart cities. The data-centric approach is embedded in a specialized digital twin focusing on the Salvador Historic Center - Pelourinho, a highly important cultural asset and tourism spot for the city of Salvador. Currently, Pelourinho faces a persistent challenge in sustaining tourism flux while safeguarding its cultural and heritage assets. Preliminary results indicate that the data-centric approach adopted by Pelourinho's DT facilitates data visualization, integrates data silos, and adequately supports management, enabling managers to address heritage preservation and conservation issues, control over-tourism, and implement urban resilience and climate adaptation measures.

\end{abstract}



\begin{teaserfigure}
   \centering
   \includegraphics[width=0.35\textwidth]{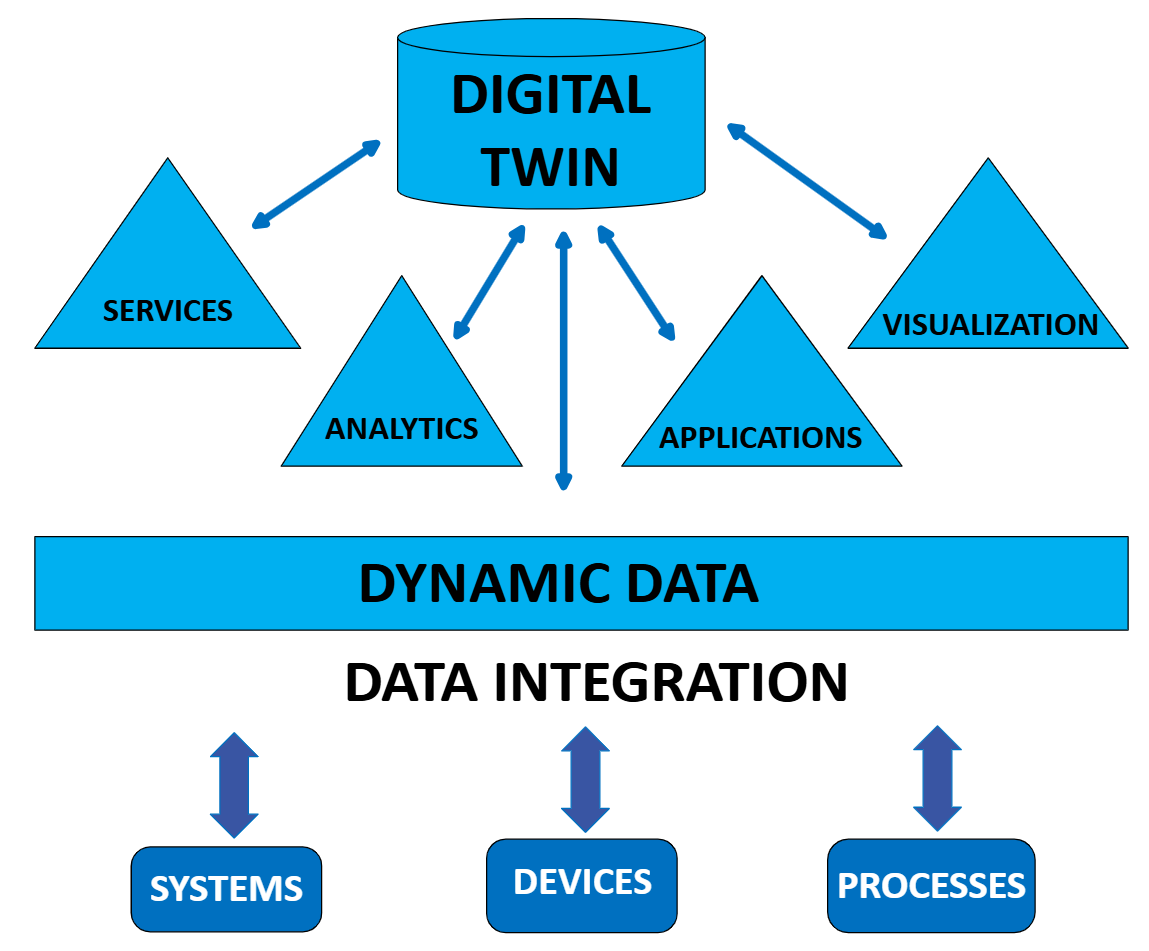}
  \caption{TEASER - Dynamic and Customized Map-based Data-centric approach for a Specialized Tourism and Cultural Heritage Digital Twins}
  \Description{}
  \label{fig:teaser}
\end{teaserfigure}

\received{20 December 2025}
\received[revised]{30 January 2026}
\received[accepted]{12 February 2026}

\maketitle

\section{Introduction}\label{sec:intro}

Tourism is an important economic activity with a huge impact on economies around the world (\cite{world_travel_and_tourism_council_travel_2025}). As discussed in (\cite{santos_evidence_2025}), 1.4 billion international tourist arrivals were recorded in 2024, generating 10.9 trillion dollars in economic activity, an amount equivalent to 10\% of the Global GDP (Gross Domestic Product). In addition, tourism activity created jobs for 357 million people, equivalent to 10.6\% of the global workforce (\cite{world_travel_and_tourism_council_travel_2025}).

Tourism global indicators suggest that tourism activity is increasing and is commonly associated with impacts on cultural heritage and the well-being of inhabitants (\cite{santos-junior_residents_2020}).

Cultural heritage preservation is a fundamental aspect of cities, including new smart cities, and digital transformation management strategies (\cite{martins_cidade_2024}). In this context, Digital Twins (DTs) can be used for tourism management and to support the preservation of cultural heritage and historic sites in general (e.g., \cite{serbouti_digital_2025}; \cite{dang_digital_2023}; \cite{akyol_digital_2025} \cite{de_souza_cidades_2025}). In summary, digital twins customized for cultural heritage create virtual replicas of historic sites that support in situ management and preservation initiatives.

\begin{figure}[b]
  \includegraphics[width=0.5\textwidth]{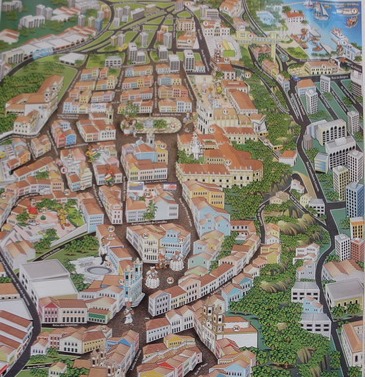}
  \caption{Salvador Historic Center (Pelourinho) (Source: \cite{bacelar_guia_2025}).}
  \Description{Smart City.}
  \label{fig:Pelourinho}
\end{figure}

The Historic Center of Salvador, Brazil, also known as Pelourinho (Figure \ref{fig:Pelourinho}), is recognized as a UNESCO World Heritage Site. It faces significant challenges related to planning, climate change adaptation, optimization, and the facilitation of tourist flux, as well as urban heat islands. These issues are exacerbated by the lack of management of tourism flux, the scarcity of green spaces, the high density of buildings, and inadequate land use, which compromise not only the environment but also the region's historical and cultural heritage.

Artificial intelligence (AI) is another important component explored in current approaches for tourism management and cultural heritage \cite{sanchez-martin_artificial_2025} \cite{santos_evidence_2025}. AI support requires a data-rich setup, in which the quantity and quality of the data are paramount for providing effective AI-based solutions (\cite{ercim_ai_2025}; \cite{sanchez-martin_artificial_2025}; \cite{almeida_digital_2025}).

The research gap in the scenario with DTs being used to manage tourism and cultural heritage is therefore threefold:
\begin{itemize}
    \item There is a gap of solutions based on digital twins (DTs) that consider tourism, cultural heritage, and social issues concomitantly;
    \item Digital twins for cultural heritage are mostly focused on simple digitization and visualization tools, and lack incorporating data-rich models; and
    \item There is a gap for digital twins (DTs) that embed artificial intelligence tools to address smart city multi-faceted issues (and related datasets), such as tourism, social well-being, and cultural heritage.
\end{itemize}


As such, the main research question addressed in this paper is as follows:

\begin{itemize}
    \item Is it possible to integrate a data-rich strategy into a specialized digital twin framework that holistically deals with tourism and cultural heritage urban issues?
\end{itemize}

The research gap in this scenario is how to incorporate several data-rich datasets from different domains and support new strategies using DT technology for urban development and smart cities.

This paper's objective is to propose a data-rich strategy integrated into a specialized QGIS-based digital twin to support integrated tourism and cultural heritage management actions in the Pelourinho Historic Center.

The paper addresses the following contributions:
\begin{itemize}
     \item Develop a data-centric approach for a new type of specialized digital twin; 
     \item Integrate IoT sensors and predictive analytics with artificial intelligence to address the impact of tourism and climate change on the preservation of the historical heritage of Pelourinho; and
    \item Develop a specialized DT enabling real-time site monitoring, scenario simulations, and predictive maintenance, offering a replicable, data-centric approach to tourism and heritage preservation.
\end{itemize}

The paper innovates by adopting QGIS maps georeferencing and providing the necessary flexibility to incorporate, on demand, dynamic data with various structural elements such as Internet of Things (IoT) sensors, heritage, historic, and social data related to the real-world scenario of historical centers, while capturing and exporting dynamic data for these elements. Practical examples of incorporated structural components include cameras, on-the-fly people and tourism flux, vehicle flux, temperature and humidity sensors, cultural and historic data elements, among others.

This paper is organized as follows: the introduction section \ref{sec:intro} presents tourism and cultural heritage preservation using DT technology. Section \ref{sec:DT} introduces the digital twins in the context of tourism and cultural heritage. Section \ref{sec:DT_Data_Centric} presents the data-centric approach used by the digital twin, followed by a use case on tourism flux and a discussion on its embedding within the digital twin architecture. Finally, Section \ref{sec:Conclusion} concludes the discussion with final considerations on the DT data-rich approach and future work.

\section{Digital Twins for Tourism and Cultural Heritage Preservation in Smart Cities} 
\label{sec:DT}

The smart city strategy is an "umbrella" trend that considers various pillars for urban development, such as mobility, infrastructure, security, health, and energy, to mention some (\cite{shao_sustainable_2025}; \cite{farid_smart_2021a}). Smart city strategies aim to promote the well-being of city inhabitants while simultaneously optimizing urban management and services.

Technology and data form the core of smart city project development.  The Internet of Things (IoT), artificial intelligence, and digital twins, to name a few, are fundamental technologies of smart city development across nearly all areas, including tourism and cultural heritage preservation (\cite{menaguale_digital_2023}).

\subsection{Digital Twins}
\label{sec:DT_Architecture}

A digital twin is a digital representation of a physical asset, process, or system that accurately replicates its data, behavior, and interaction with other assets. DTs enable real-time monitoring, situation simulation, and data analysis, providing valuable insights into the performance and behavior of the modeled counterparts (\cite{mazzetto_review_2024}; \cite{afif_supianto_urban_2024}).

Digital Twins emerge as a promising tool for urban development. They allow virtual modeling of the urban environment and enable real-time data analysis to support tailored strategies and policies for smart, sustainable, and resilient cities (\cite{afif_supianto_urban_2024}).

Digital twins for tourism and cultural heritage preservation are a current trend (\cite{almeida_digital_2025}). A specialized digital twin for tourism and cultural heritage that handles dynamic data and integrates data across different applications is shown in Figure \ref{fig:teaser}. One fundamental aspect of this proposal is the data-rich approach, which is pivotal for tourism and cultural heritage and relies heavily on artificial intelligence.

\subsection{Digital Twin for Tourism and Heritage Preservation - The Pelourinho Historic Center Issues}
\label{sec:DT_Architecture}

The Historic Center of Salvador is one of the most significant cultural and architectural heritage sites in Brazil and was designated a UNESCO World Heritage Site. Despite its historical, symbolic, and economic significance, Pelourinho faces a set of persistent, interrelated urban challenges that directly affect both tourism sustainability and heritage preservation. These challenges are intensified by the area's complex spatial configuration, its high tourist appeal, and the limitations of traditional urban management approaches.

One of the central issues concerns the lack of systematic management of tourist flows. Tourism activities in Pelourinho are highly concentrated in specific streets, squares, and time periods, leading to spatial and temporal imbalances. The absence of continuous monitoring mechanisms results in overcrowding during peak hours and underutilization of other areas, increasing pressure on historic buildings, public spaces, and urban infrastructure. This phenomenon contributes to overtourism, reducing the quality of the visitor experience and negatively impacting residents’ well-being.

Another critical problem concerns human density and congestion in heritage-sensitive spaces. Excessive visitor concentrations accelerate physical degradation processes, such as the wear of pavements, facades, and public equipment, while also increasing risks to safety, accessibility, and emergency response. The lack of real-time data and predictive tools limits public authorities' capacity to anticipate and mitigate these effects.

Pelourinho also faces significant environmental challenges, particularly those related to urban heat and noise pollution. The high density of built structures, limited vegetation coverage, and intense pedestrian activity contribute to the formation of urban heat islands, which affect thermal comfort for both residents and tourists. In parallel, cultural events, street performances, vehicular circulation in surrounding areas, and large tourist crowds generate elevated noise levels, compromising the acoustic comfort of humans and animals and potentially affecting the structural integrity of historic buildings over time.

In addition to these issues, data fragmentation and institutional silos constitute structural limitations to integrated urban management. Information on tourism, cultural heritage, environmental monitoring, and urban infrastructure is typically dispersed across multiple institutions and formats, hindering comprehensive analysis and coordinated decision-making. This fragmentation restricts the ability to establish correlations between tourism dynamics, environmental conditions, and heritage conservation indicators.

Furthermore, Pelourinho is increasingly exposed to climate change-related risks, including rising temperatures and more frequent extreme weather events. These factors exacerbate existing vulnerabilities of historic structures and public spaces, demanding adaptive and resilient management strategies supported by data-driven tools.

Given this context, Pelourinho constitutes a complex urban environment in which tourism pressure, environmental stress, and challenges of heritage preservation coexist and interact. Addressing these issues requires an integrated approach that combines real-time monitoring, historical data, spatial analysis, and predictive modeling. In this sense, a specialized Digital Twin emerges as a suitable technological solution to support the analysis, simulation, and management of tourist flows, environmental impacts, and preservation strategies in the Historic Center. By centralizing heterogeneous data and enabling dynamic representations of urban processes, the Pelourinho Digital Twin aims to provide actionable insights for sustainable tourism management and the long-term preservation of cultural heritage.

\begin{figure*}[!t]
\centering
\includegraphics[width=0.8\textwidth]{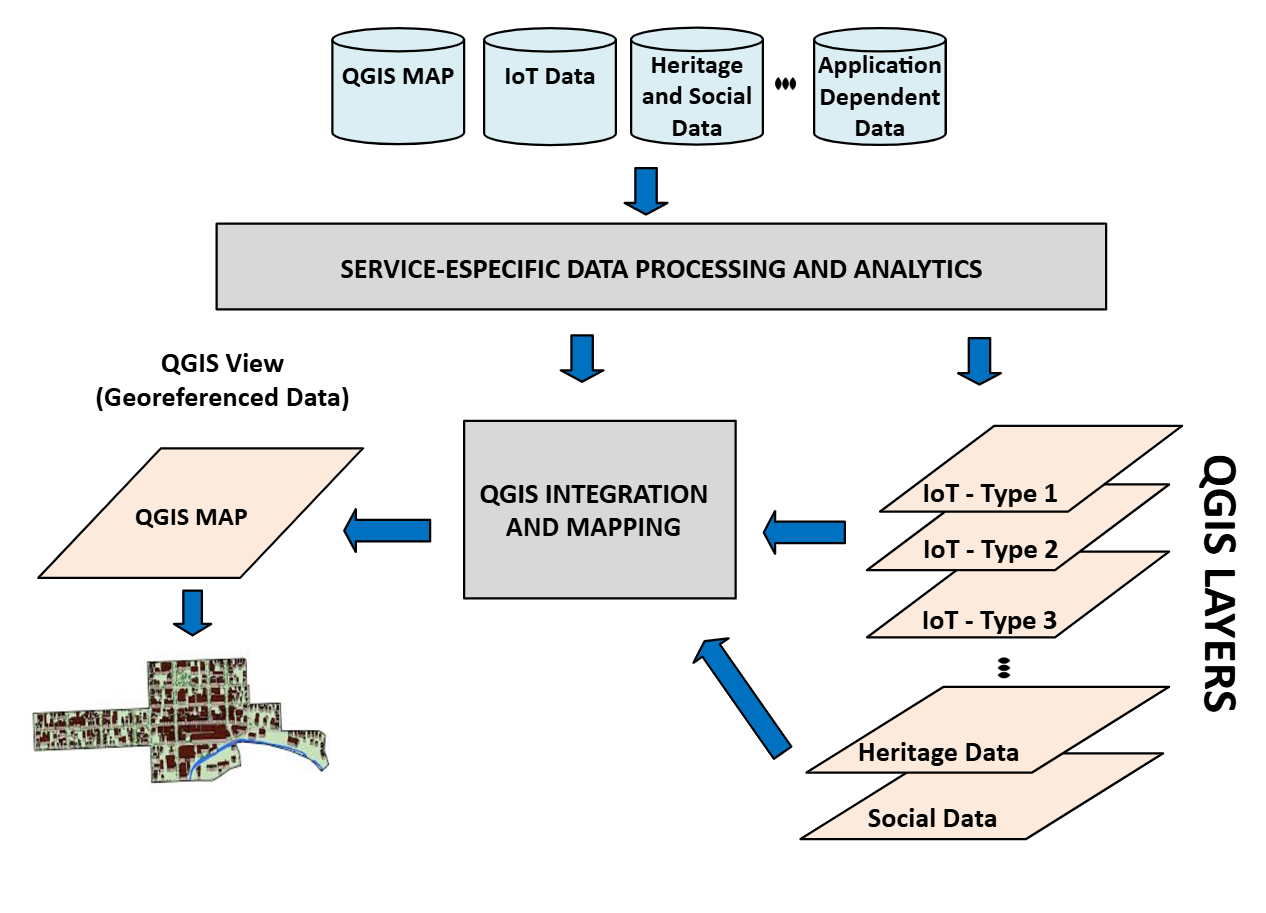}
\caption{Pelourinho's DT Data-Centric Approach.}
\label{fig:datacentric}
\end{figure*}

\section{Digital Twin with Dynamic and Customized Data-Centric and QGIS Map-based Approach} \label{sec:DT_Data_Centric}

The QGIS-based data-centric approach of the digital twin of Pelourinho (Pelourinho's DT) (Figure \ref{fig:datacentric}) provides several key features for managers, urban developers, and urban planners within the context of a smart city strategy. The supported key features are:
\begin{itemize}
    \item The integration of heterogeneous and diverse data sources from IoT physical and logical sensors, social data from institutions dealing with Pelourinho's settlement, and historical and cultural heritage data elements;
    \item Flexible data analytics processing with multiple artificial intelligence tools and algorithms; and 
    \item Use of georeferenced data to facilitate spatial-data correlation and management visualization with dynamic and flexible input parameters.
\end{itemize}

\subsection{The Pelourinho's DT Databases and Data-Centric Approach}

The main principle of the data-centric approach adopted in this research is to shift from advanced data models toward improving the quality and quantity of the data. One of the main arguments for this approach comes from the fact that Pelourinho's DT uses artificial intelligence, and AI is moving to data-centric AI since the algorithms need data with quantity and quality (\cite{zha_data-centric_2025}; \cite{siegl_data-centric_2016}).

The Pelourinho's DT data-centric requirements are as follows:

\begin{itemize}
    \item Dynamically integrate time-series data acquired with the help of various types of IoT sensors for distinct DT's situational analysis;
    \item Support diverse and specific types of data necessary for analyzing the situation of Pelourinho's tourism, cultural heritage, and climate change; and
    \item To georeference all data relative to the QGIS map to support real-location in Pelourinho's in-site management and impact analysis.
\end{itemize}

The Pelourinho's database is composed of the following elements (Figure \ref{fig:datacentric}):
\begin{itemize}
    \item A geographic QGIS database;
    \item An IoT-based database of sensed parameters relevant for Pelourinho's tourism and climate change impact analysis;
    \item A heritage-related historical database and social data database; and
    \item An application-defined database for supporting Pelourinho's DT customization for different smart city management strategies and applications.
\end{itemize}

The geographic and georeferenced QGIS database stores all data in accordance with QGIS map-processing requirements and operations.

The IoT-based database stores all physical and sensor parameters associated with the services and applications supported by the DT, including tourism and climate change applications. All data is time-referenced to allow time-series analysis supported by the DT.

The application-specific database includes data parameters, knowledge, and information concerning the custom application and services supported by the DT.

QGIS does not explicitly support a dynamic, data-centric approach. As such, the Pelourinho's DT employs a database-to-QGIS mapping approach, as illustrated in Figure \ref{fig:datacentric}. In summary, it works as follows:
\begin{itemize}
    \item The used database stores IoT and other types of data, allowing a generic set of service-specific data processing and data analysis by distinct AI algorithms;
    \item Data stored in the databases map to the layers' structure used by the QGIS maps; and 
    \item The deployed database structure ensures item Time-series mappings and relations.
\end{itemize}

\subsubsection{Pelourinho's DT Data-Centric Deployment}

The deployment of Pelourinho's Digital Twin adopts a data-centric approach, in which data serves as the architecture's structuring element. At the same time, applications and visualization mechanisms act as access and analysis layers.

The database used in Pelourinho's DT is the open-source PostgreSQL with the PostGIS extension, which supports spatial data storage and manipulation (\cite{salunke_performance_2024}).

A PostgreSQL database was chosen due to QGIS's support for PostGIS, including spatial capabilities for 2D, 3D, and 4D, spatial relationships, and spatial analytics. In addition, it is high-performance with large datasets, ensuring the DT's scalability for large-scale application scenarios.

As illustrated in Figure \ref{fig:datacentric}, the overall data deployment is based on a set of logically distinct yet integrated databases (IoT, cultural heritage, social, and application-dependent data) that feed a Service-Specific Data Processing and Analytics (SSDPA) module and, subsequently, the QGIS-based integration and visualization layer.

The service-specific data processing and analytics (SSDPA) module is part of the DT main architecture. Regarding data manipulation, the SSDPA supports spatial SQL queries and database-defined views.

The main characteristics of the data-centric deployment adopted in the Pelourinho's DT are as follows:

\begin{itemize}
    \item A central spatial database stores different types of data;
    \item Service-specific pre-processing is performed directly on the database using spatial SQL;
    \item QGIS acts as an integration and visualization layer; and
    \item The database tables and views reflect the QGIS layer structure used for georeferenced operations. 
\end{itemize}

\subsection{A Use Case: Pelourinho's Tourism Flux}

\begin{figure*}[t]
  \includegraphics[width=1\textwidth]{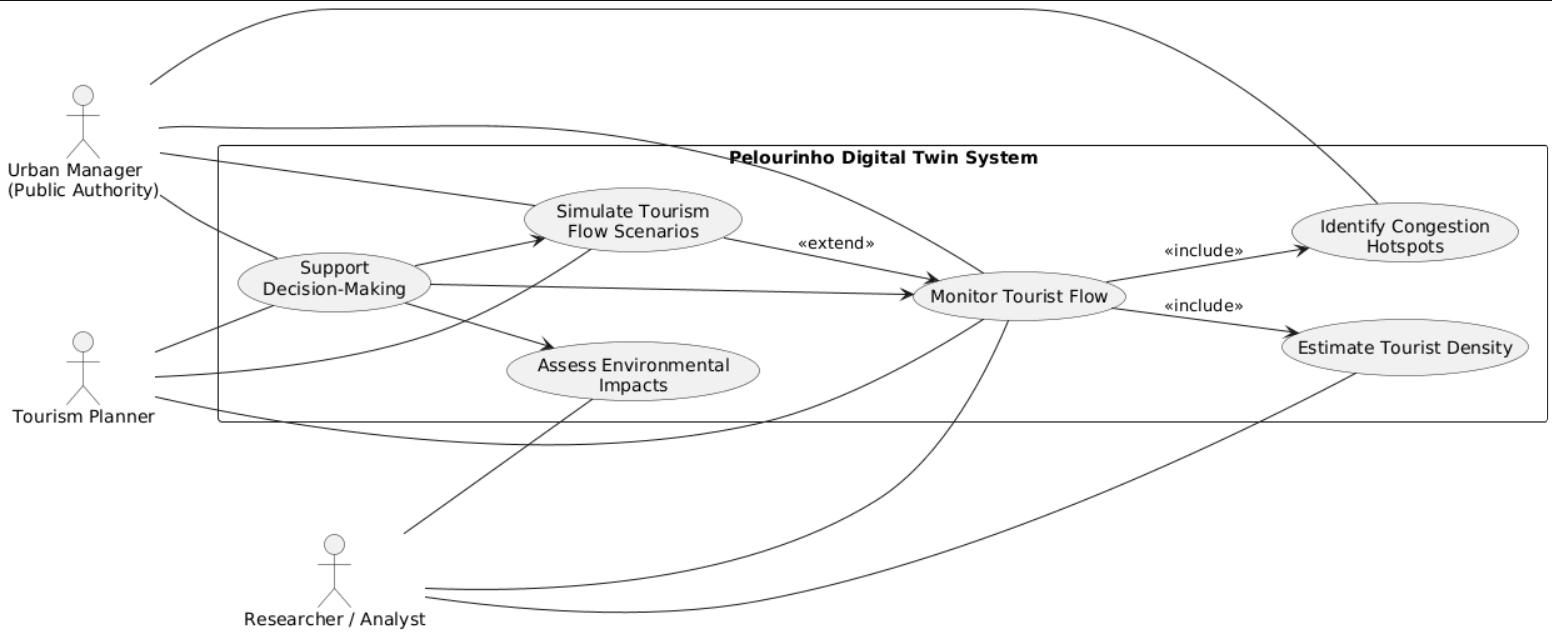}
  \caption{Case Diagram - Pelourinho's DT Tourism Flux Service}
  \Description{}
  \label{fig:DiagramaUC.png}
\end{figure*}

The use case diagram presented in this study formalizes a practical application of the Pelourinho Digital Twin focused on the analysis and management of tourist flux within the Historic Center of Salvador. The diagram defines the main actors, the system boundaries, and the interactions that enable the Digital Twin to operate as a monitoring, analytical, and decision-support platform.

Three primary actors interact with the Pelourinho Digital Twin System. The Urban Manager (Public Authority) represents municipal and heritage management institutions responsible for regulating urban space, tourism activities, and preservation policies. The Tourism Planner acts as a strategic user focused on organizing events, designing tourist routes, and managing visitation patterns. The Researcher/Analyst represents academic and technical users who employ the Digital Twin to investigate spatial, temporal, and environmental dynamics associated with tourism.

At the core of the diagram lies the Pelourinho Digital Twin System, which encapsulates the data-centric infrastructure, analytical models, and visualization tools described in the architectural framework. The central use case, Monitor Tourist Flow, constitutes the system's primary functionality. It enables stakeholders to visualize and track the spatial and temporal distribution of tourists across streets, squares, and points of interest, using dynamic maps and indicators generated from integrated data sources.

This core functionality explicitly includes two subordinate use cases: Estimate Tourist Density and Identify Congestion Hotspots. The first relies on the fusion of heterogeneous data collected from different sources, such as camera detections, passive WiFi/Bluetooth signals, and statistical estimates, to generate reliable density measures. The second use case builds on these estimates to detect areas of excessive concentration, thereby supporting early identification of overcrowding risks in heritage-sensitive spaces.

The use case Simulate Tourism Flow Scenarios extends the monitoring process by enabling prospective analyses. Through scenario simulation, stakeholders can evaluate “what-if” situations, such as cultural events, changes in pedestrian routes, or seasonal variations in visitation. This function highlights the predictive capacity of the Digital Twin, distinguishing it from conventional monitoring systems.

Another relevant use case, Assess Environmental Impacts, connects tourist flow patterns with environmental variables, particularly urban noise levels and microclimatic conditions. By correlating human density with acoustic pressure and temperature, the Digital Twin supports integrated assessments of comfort, sustainability, and heritage preservation.

Finally, the use case Support Decision-Making aggregates the outputs of monitoring, estimation, simulation, and environmental assessment. It represents the ultimate purpose of the Digital Twin: to provide structured, data-driven insights that inform urban governance, tourism planning, and preservation strategies. This use case depends on the results of multiple analytical processes, reinforcing the systemic and integrative nature of the proposed Digital Twin.

Overall, the use case diagram demonstrates how the Pelourinho Digital Twin uses a data-centric approach to tourism management. By clearly defining actors, system functions, and their relationships, the diagram illustrates how heterogeneous data are transformed into actionable knowledge, supporting both real-time management and strategic planning in a complex historic urban environment. The use case diagram can be identified below

.

\subsection{The Data-Centric Approach embedded in the Pelourinho's DT Architecture}

\begin{figure*}[bt]
  \includegraphics[width=0.7\textwidth]{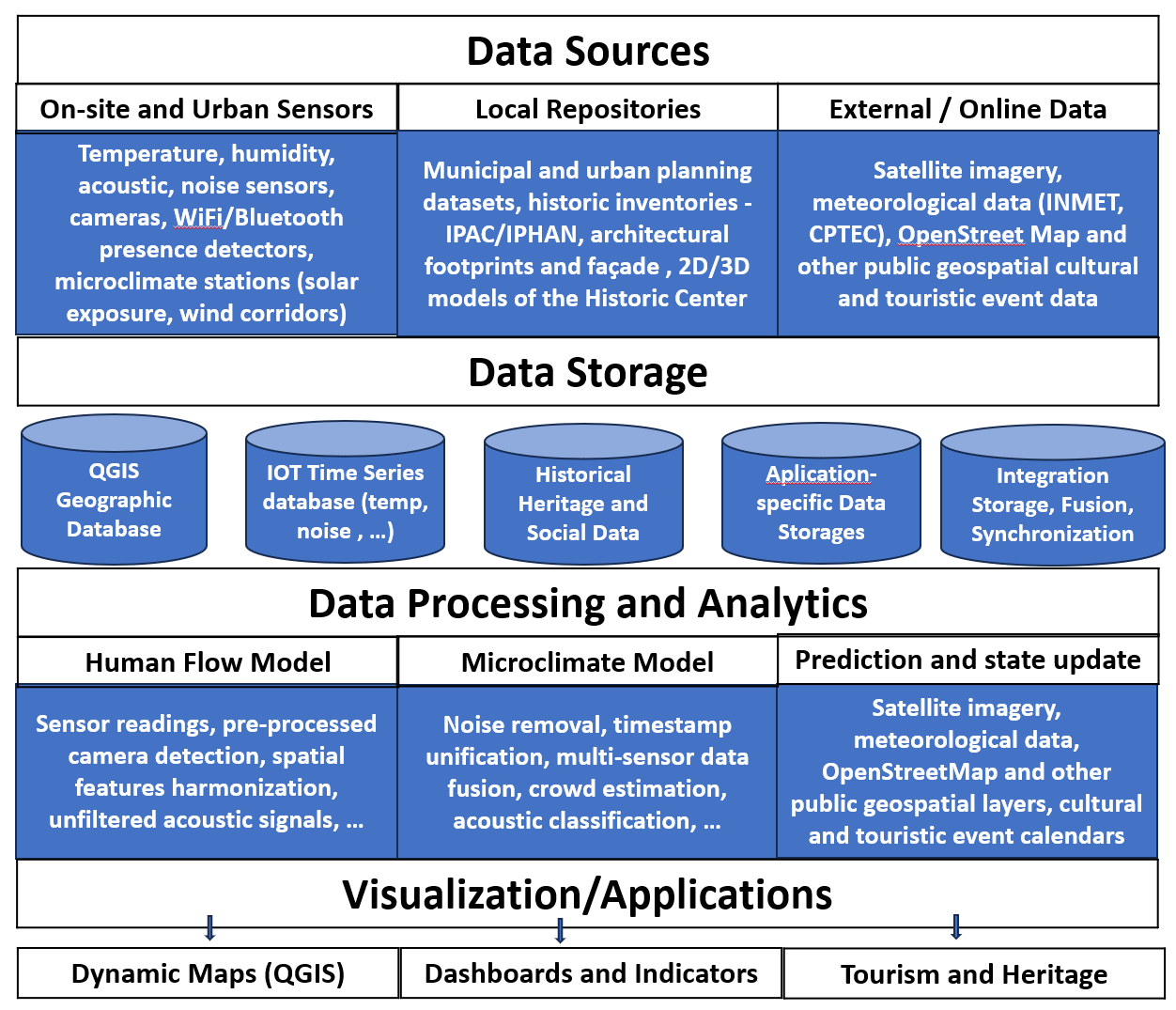}
  \caption{Pelourinho's Digital Twin Architecture Layers}
  \Description{}
  \label{fig:PelourinhoDT_Architecture}
\end{figure*}

The Pelourinho's digital twin architecture is illustrated in Figure \ref{fig:PelourinhoDT_Architecture}. It is composed of five layers, applications, and components (\cite{de_souza_cidades_2025}). The architecture's layers are hierarchical and interact together to represent, monitor, and optimize the physical environment. The physical layer represents the real world, including physical assets (buildings, urban infrastructure, equipment, vehicles, people, and social-related data, among others).

The proposed architecture for the Pelourinho Digital Twin was developed from a systemic vision that integrates heterogeneous data sources, a centralized processing core, and analytical models for simulating urban phenomena. The figure developed in the research's conceptual model illustrates this organization by presenting, in a hierarchical manner, the elements comprising the physical environment, the data flows, and the computational layers that support the operation of the digital twin.

The architecture is structured in three main blocks. The first layer corresponds to the real world, where sensors and devices distributed throughout the Historic Center capture essential variables such as temperature, humidity, noise level, people flow, and microclimatic conditions. This data is complemented by institutional digital sources, such as the urban registry; heritage inventories from IPAC (\textit{Instituto do Patrimônio Artístico e Cultural da Bahia})\footnote{IPAC and IPHAN are governmental institutions in Bahia and Brazil that lead Cultural Heritage across the state and country.} and IPHAN (\textit{Instituto do Patrimônio Histórico e Artístico Nacional na Bahia}); and existing 2D/3D models, as well as external resources, including satellite images, national meteorological data, and public geospatial databases. This layer is characterized by the diversity of formats, resolutions, and periodicities, which reinforces the need for a robust integration mechanism.

The second layer represents precisely this mechanism: the Digital Twin Data Hub, which constitutes the core of the data-centric strategy. Inspired by the logic presented in data-centric architectures for heritage preservation, the Pelourinho Data Hub integrates three essential components:

• The raw data zone, which directly receives sensor records, camera detections, acoustic data, and pre-processed spatial features;

• The processing and harmonization pipeline, responsible for noise removal, inconsistency correction, temporal synchronization, georeferencing via QGIS base layers, and merging multiple sources, especially for estimating population density;

• The structured repositories, composed of geographic databases, time series databases, and application-specific datastores. This set is the organizing element of the entire architecture, as it ensures spatial, temporal, and semantic coherence to the data used by the digital twin models.

The third layer comprises analytical models and the simulation core, in which structured data are transformed into dynamic representations of the territory. At this level, human flow, microclimate, and sound-propagation models operate, using the Data Hub's geospatial and temporal databases to generate maps, indicators, forecasts, and simulated scenarios. This layer also includes predictive analytics modules and continuous updating mechanisms, ensuring that the digital twin reflects the most recent state of Pelourinho.

Finally, the top layer of visualization and applications provides dynamic maps, dashboards, and interfaces that support urban monitoring, cultural heritage management, and planned intervention in public space.

This organization demonstrates that the data-centric strategy is not an additional element, but the structural axis that articulates all the components of the architecture. While the real world provides diverse and fragmented data, and analytical models require consistent databases to function, it is in the Data Hub that the critical mediation between these two dimensions occurs.

The overall architecture is sustainable only because the data-centric strategy ensures that data is received, structured, spatialized, and integrated, enabling the digital twin to operate with precision, continuous updates, and simulation capabilities. In summary, the presented proposal encompasses all the necessary elements: sensors, institutional databases, processing, models, and applications, and illustrates how data centralization is the fundamental component that sustains the Pelourinho Digital Twin.

\section{Final Considerations} \label{sec:Conclusion}

Specialized digital twins are a trend in cultural heritage and tourism management in smart cities and digital transformation contexts, where over-tourism and the preservation of cultural heritage assets must be planned and enforced.

A data-rich approach is a must for specialized digital twins. The QGIS map-based data-centric approach developed for Pelourinho's DT enables the integration of heterogeneous data—spatial, temporal, and semantic—into a central repository, ensuring consistency, scalability, and flexibility for diverse uses of the Digital Twin.

In the case of Salvador Historic Center (Pelourinho), the data-rich approach embedded in the Pelourinho's DT architecture facilitated the management and decision-making processes of managers involved in the tourism economy and in the preservation of cultural heritage assets, activities that are typically distributed over diverse institutions.

Future work includes developing a new set of services and applications for the Salvador Historic Center (Pelourinho) for tourism and cultural heritage, and mitigating the impact of social and cultural events in Pelourinho.

In the context of tourism, we have examples such as urban heat-island modeling, climate-risk scenarios, real-time visitor flow monitoring, and dynamic route recommendation systems. In the context of cultural heritage and monument preservation, we have examples such as façade degradation forecasting based on heat stress, digital inventory of heritage assets, simulation of restoration scenarios, and green areas planning and deployment.



\begin{acks}

The authors thank the ANIMA Institute for scholarship support 2025/2026.

\end{acks}




\bibliographystyle{ACM-Reference-Format}
\bibliography{ADVANCE} 

\end{document}